\documentclass[secnumarabic, graphics,floatfix, nofootinbib,tightenlines,nobibnotes, aps, prl, 12pt]{revtex4-2}
\usepackage[english]{babel}
\usepackage[utf8]{inputenc}
\usepackage{amsfonts}
\usepackage{multirow}
\usepackage{graphicx}
\usepackage{epstopdf}
\usepackage{hyperref}
\usepackage{latexsym}
\usepackage{amsmath}
\usepackage{amssymb}

\usepackage[utf8]{inputenc}
\hypersetup{
    colorlinks=true,
    linkcolor=red,
    citecolor=blue,
}
\usepackage{color}
\usepackage[T1]{fontenc}
\usepackage{txfonts}
\usepackage{mathrsfs}
\usepackage[center]{subfigure}

\begin{document}
 \setcounter{secnumdepth}{2}
 \newcommand{\bq}{\begin{equation}}
 \newcommand{\eq}{\end{equation}}
 \newcommand{\bqn}{\begin{eqnarray}}
 \newcommand{\eqn}{\end{eqnarray}}
 \newcommand{\nb}{\nonumber}
 \newcommand{\lb}{\label}
 
\title{Parameter estimation for Einstein-dilaton-Gauss-Bonnet gravity with ringdown signals}

\author{Cai-Ying Shao$^{1}$}
\author{Yu Hu$^{1}$}\email[E-mail: ]{yuhu@hust.edu.cn}
\author{Cheng-Gang Shao$^{1}$}\email[E-mail: ]{cgshao@hust.edu.cn}

\affiliation{$^{1}$ MOE Key Laboratory of Fundamental Physical Quantities Measurement, Hubei Key Laboratory of Gravitation and Quantum Physics, PGMF, and School of Physics, Huazhong University of Science and Technology, 430074, Wuhan, Hubei, China}

\date{\today}

\begin{abstract}
Future space-based gravitational-wave detectors will detect gravitational waves with high sensitivity in the millihertz frequency band, which provides more opportunities to test theories of gravity than ground-based ones.
The study of quasinormal modes (QNMs) and their application to testing gravity theories have been an important aspect in the field of gravitational physics.
In this study, we investigate the capability of future space-based gravitational wave detectors such as LISA, TaiJi, and TianQin to constrain the dimensionless deviating parameter for Einstein-dilaton-Gauss-Bonnet (EdGB) gravity with ringdown signals from the merger of binary black holes.
The ringdown signal is modeled by the two strongest QNMs in EdGB gravity. 
Taking into account time-delay interferometry, we calculate the signal-to-noise ratio (SNR) of different space-based detectors for ringdown signals to analyze their capabilities.
The Fisher information matrix is employed to analyze the accuracy of parameter estimation, with particular focus on the dimensionless deviating parameter for EdGB gravity. 
The impact of the parameters of gravitational wave sources on the estimation accuracy of the dimensionless deviating parameter has also been studied.
We find that the constraint ability of EdGB gravity is limited because the uncertainty of the dimensionless deviating parameter increases with the decrease of the dimensionless deviating parameter.
LISA and TaiJi has more advantages to constrain the dimensionless deviating parameter to a more accurate level for the massive black hole, while TianQin is more suitable for less massive black holes.
Bayesian inference method is used to perform parameter estimation on simulated data, which verifies the reliability of the conclusion.
\end{abstract}

\maketitle

\newpage
\section{Introduction}

Gravitational wave signals from the coalescence of compact binaries were detected by LIGO Scientific Collaboration and Virgo Collaboration~\cite{prl-GW150914} for the first time in 2015.
This landmark detection has ushered in a new frontier of testing General Relativity and exploring our universe~\cite{prl-tests-gr,prl-constraints-dipole,prd-theoretical-physics,cqg-bh-gws,grg-extreme-gravity-tests,grg-extreme-gravity}.
One of the promising prospects is analyzing the QNMs in a ringdown waveform.
These oscillation frequencies are determined by the mass and angular momentum of the remnant black hole.
Empirical detection of QNMs would not only provide us an opportunity to test the no-hair theorem~\cite{prl-no-hair-GW150914,prd-bayesian-model,prd-science-tianqin,prd-no-hair-prospect} but also allow to constrain the alternative theory~\cite{prd-modified-constrain,prd-edgb-prob,prd-nonsingular,prd-gravitational-chern-simons}.
To this end, the technological conditions for larger SNR are required.
The ground-based gravitational wave detectors (such as LIGO, Virgo, and KAGRA) operate in frequency bands above 10Hz and suffer the influence of the gravity gradient and seismic noise, which leads to the ringdown signal being neglected for short durations.
For proposed space-based gravitational wave detectors, the frequency bands dip to the millihertz frequency band, where more plentiful sources are waiting to be probed. 
It is possible for future space-based observations such as LISA, TaiJi, and TianQin to detect ringdown signals from intermediate and supermassive sources with rather large SNR.
On this basis, the window for testing the nature of gravity will be opened, which is significant.

Einstein’s theory of general relativity, as the cornerstone of modern gravitational theories, describes the most beautiful physical world at macroscopic scales.
However, such a prevailing theory is plugged with the black hole singularity, dark matter and dark energy, and the non-renormalization problems, which indicate that general relativity is not a complete theory of gravity.
A variety of conceiving alternative theories of gravity have been proposed owing to complexities of curvature in General Relativity such as Lovelock~\cite{jmp-einstein-gener}, $f(R)$~\cite{lrr-f} and $f(T)$~\cite{rr-gauge-theory} theories.
As a higher-curvature gravity theory motivated by low-energy limit in string gravity~\cite{cqg-higher-derivative-string}, EdGB gravity was formulated that a dilaton scalar field is coupled to the Gauss-Bonnet invariant with coupling parameter ${{\alpha _{{\rm{GB}}}}}$ in the action~\cite{prd-charg-string,prd-dilatonic}.
As a consequence, the field equations are always of second order and this gravity is ghost-free.
Furthermore, the black hole and neutron star can become scalarized spontaneously in this theory~\cite{prl-curvature-scalarization,prl-spontaneous-scalarization,prd-scalarized-black-hole}.
The presence of a nontrivial scalar field outside its horizon leads to the violation of the classical ``no-hair'' theorems~\cite{prd-dilatonic-higher,prl-gr-attractor} and thus the resultant hairy black hole can not be described by the Kerr hypothesis.
It is exactly the fact that there are a number of attractive features in EdGB gravity that have led to observable implications~\cite{Cunha:2016wzk,prd-theoretical-physics}.

At present, with the development and gradual maturity of technology, not only is EdGB gravity constrained by astronomical observations~\cite{cqg-test-gr-astrophysical-observations,n-radio-links-cassini,prd-constraint-low-mass-X} but also by gravitational wave data from binary black holes~\cite{prl-fundamental-physics,prd-test-amplitude-corrections,prd-constraints-higher-order-curvature} and numerical relativity simulation~\cite{prd-num-relativity-simulation,prd-binary-mergers}.
Here, inspired from the significance of quasinormal mode corrections for a massive black hole in EdGB gravity~\cite{prd-ringdown,prd-edgb-prob} and high enough SNR with space-based gravitational wave detectors, we investigate the ability to estimate accuracy of the dimensionless deviating parameter with LISA, TaiJi and TianQin in parameterized ringdown signal.
According to the standard dimensional analysis, it seems the dimensionful coupling parameter ${{\alpha _{{\rm{GB}}}}}$ leads to negligible deviations in low energy regime, equivalently at the scale of supermassive black holes.
However, in lack of a complete quantum theory of gravity, it is better to resort to the experiment data to constrain the deviation to general relativity independently at the supermassive black hole scale and the stellar-origin black hole scale.
With this in mind, we focus mainly on space-based gravitational wave detectors, with the supermassive black hole as target.
First, we build the ringdown waveform with the two strongest QNMs in EdGB gravity.
The black hole in question is extremely slowly-rotating with first order in rotation, whose QNMs in gravitational perturbations naturally return to Kerr spacetime as the dimensionless deviating parameter fades away~\cite{prd-qnm-edgb-first}.
In order to further improve the SNR, it is essential to eliminate different noise.
The laser frequency noise in an unequal-arm interferometer is not negligible~\cite{prd-cancellation}.
To suppress laser noise, the time-delay interferometry combination has been proposed~\cite{tinto2021time}.
In this paper, we choose the first-generation TDI Michelson combination X~\cite{tinto2021time} to obtain response functions and noise power spectral density.
Then the SNR is calculated to assess the scientific performance of LISA, TaiJi, and TianQin.
Furthermore, the effects of the arm length of the detector, the mass, the luminosity distance, the spin of the remnant black hole, the symmetric mass ratio, and the dimensionless deviating parameter on the measurement errors of the dimensionless deviating parameter are further probed.
In particular, we present the maximum constraint on the dimensionless deviating parameter for EdGB gravity with ringdown signals from massive binary black holes.
In order to verify the conclusions, we also performed simulations of Bayesian inference to obtain probability distributions of the dimensionless deviating parameter.

The remainder of our paper is organized as follows. 
In the next section, we give a brief introduction to EdGB gravity, including the black hole solution and QNMs of gravitational field perturbation for slowly-rotating black holes.
In Section \ref{section3}, we present the ringdown signals of time-delay interferometry Michelson combination X and the sensitivity curves of space-based gravitational wave detectors.
In Section \ref{section4}, we calculate the SNR and use Fisher information matrix to analyze how measurement errors of the dimensionless deviating parameter can be affected by the arm length of the detector, the mass, the luminosity distance, the spin of the remnant black hole, the symmetric mass ratio and the dimensionless deviating parameter.
The maximum capacity of constraint on the dimensionless deviating parameter for EdGB gravity with LISA and TianQin is also presented.
Then in Section \ref{section5}, Bayesian parameter estimation is performed on simulated data to verify the reliability of the conclusion.
The concluding remarks are provided in the last section. 

\section{Einstein-dilaton-Gauss-Bonnet gravity} \lb{section2}

Let us start with the action for EdGB gravity~\cite{prd-charg-string,prd-dilatonic}
\begin{equation}\label{N1}
S = \int {{d^4}} x\frac{{\sqrt { - g} }}{{16\pi }}\left( {R - \frac{1}{2}{\partial _\mu }\phi {\partial ^\mu }\phi  + \frac{{{\alpha _{GB}}}}{4}{e^\phi }R_{GB}^2} \right) + {S_m},
\end{equation}
where ${S_m}$ is the matter sector, $g$ denotes the determinant of the metric, $R$ is the Ricci scalar, $\phi $ is a dynamical scalar field, ${{\alpha _{{\rm{GB}}}}}$ is the coupling parameter with the dimensions of a quadratic length, and ${{\cal R}_{{\rm{GB}}}^2}$ is the Gauss-Bonnet invariant expressed as
\begin{equation}\label{N2}
{\cal R}_{{\rm{GB}}}^2 = {R_{\mu \nu \rho \sigma }}{R^{\mu \nu \rho \sigma }} - 4{R_{\mu \nu }}{R^{\mu \nu }} + {R^2}.
\end{equation}
As EdGB gravity is proposed, the construction of black hole solutions has aroused considerable interest~\cite{prd-dilatonic-higher,prd-edgb,prl-rotating-edgb,prd-slowly-bh-edgb,prd-edgb-finite-coupling}.
In general, a small-coupling regime~\cite{prd-charg-string,prd-slowly-bh-edgb} is adopted to simplify the calculation as one confronts the problem of clumsy equations in the analytical solution.
For spinning EdGB black holes, the field equation can be solved by expanding the spin ${\chi _f} \ll 1$ and the dimensionless deviating parameter $\frac{{{\alpha _{{\rm{GB}}}}}}{{{M^2}}} \ll 1$ to the ideal order.
By comparison with Kerr black holes, they possess a minimal mass and larger angular momentum~\cite{prl-rotating-edgb,Blazquez-Salcedo:2016yka,prd-quadrupole,prd-spinning-bh}.

To simplify the notation, the dimensionless deviating parameter is given by
\begin{equation}\label{N3}
\zeta  = \frac{{{\alpha _{{\rm{GB}}}}}}{{{M^2}}},
\end{equation}
where $M$ is the mass of the black hole.

Furthermore, the QNMs have been studied extensively in EdGB gravity~\cite{prd-edgb-static,prd-qnm-edgb-bh,prd-qnm-edgb-first,prd-qnm-edgb-second}.
For our purpose, we only focus on the gravitational-led QNMs of extremely slowly-rotating black holes with first order in the spin. 
The resultant QNMs can be modeled as~\cite{prd-qnm-edgb-first}
\begin{equation}\label{N4qnm}
{\omega ^{nlm}}({\chi _f},\zeta ) = \omega _0^{nl}(\zeta ) + {\chi _f}m\omega _1^{nl}(\zeta ).
\end{equation}
Here, $\omega _0^{nl}$ is the QNMs of a non-rotating black hole in the polar sector~\cite{prd-edgb-static} and $\omega _1^{nl}$ is the QNMs of spin corrections.
${\chi _f}$ is dimensionless spin, ${\chi _f} = J/{M^2}$, where $J$ is the spin angular momentum of the black hole.
$m$ is the azimuthal number.
Specifically, for lowest-lying QNMs with the multipole number $l = 2,3$, the analytic fitting formula reads~\cite{prd-edgb-static,prd-qnm-edgb-first}
\begin{equation}\label{N5qnm2}
\begin{array}{l}
M\omega _0^{0l} = (1 + {f_1}{\zeta ^2} + {f_2}{\zeta ^3} + {f_3}{\zeta ^4})\omega _s^{0l},\\
M\omega _1^{0l} = {q_1} + {q_2}{\zeta ^2} + {q_3}{\zeta ^3} + {q_4}{\zeta ^4} + {q_5}{\zeta ^5} + {q_6}{\zeta ^6}.
\end{array}
\end{equation}
Here, $\omega _s^{0l}$ is the QNMs of Schwarzschild spacetime. 
For the modes with $l = 2$, $M{\omega _s^{02}} \approx 0.37370 - 0.08896i$.
For the modes with $l = 3$, $M{\omega _s^{03}} \approx 0.5994 - 0.0927i$.
The coefficients ${f_i}$ and ${q_i}$ are listed in Table~\ref{tab1}.
As the dimensionless deviating parameter tends to zero, the QNMs of EdGB gravity fall into Kerr spacetime~\cite{prd-qnm-edgb-first,prd-qnm-edgb-second,prd-ringing-rotating}.

\begin{table}[]
\caption{The coefficients for Eq.~(\ref{N5qnm2}), where ${\omega _R}$ and ${\omega _I}$ denote the real and imaginary parts of the QNMs, respectively.
The data is taken from~\cite{prd-edgb-static,prd-qnm-edgb-first}.
}
\setlength{\tabcolsep}{0.8mm}
\begin{tabular}{|c|c|c|c|c|c|c|c|c|c|c|}
\hline
$(l,m)$                 & QNMs &  ${f_1}$       & ${f_2}$       & ${f_3}$      & ${q_1}$      & ${q_2}$      & ${q_3}$       & ${q_4}$       & ${q_5}$      & ${q_6}$      \\ \hline
\multirow{2}{*}{(2,2)} & ${\omega _R}$   & -0.03135 & -0.09674 & 0.23750 & 0.06290  & -0.01560 & -0.00758 & -0.06440  & 0.26800   & -0.60300  \\ \cline{2-11} 
                    & ${\omega _I}$    &  0.04371  & 0.17940   & -0.29470 & 0.00099 & -0.00110 & 0.01864  & -0.17271 & 0.56422 & -0.81190 \\ \hline
\multirow{2}{*}{(3,3)} & ${\omega _R}$    &  -0.09911 & -0.04907 & 0.09286 & 0.06740  & -0.02910 & 0.02510   & -0.32090  & 1.17030  & -1.33410 \\ \cline{2-11} 
                    & ${\omega _I}$    &  0.07710  & 0.13990   & -0.34500  & 0.00065 & 0.00023 & 0.02330  & -0.28320  & 1.32300   & -2.44200  \\ \hline
\end{tabular}
\label{tab1}
\end{table}

\begin{figure}[]
\centering
\includegraphics[scale=0.45]{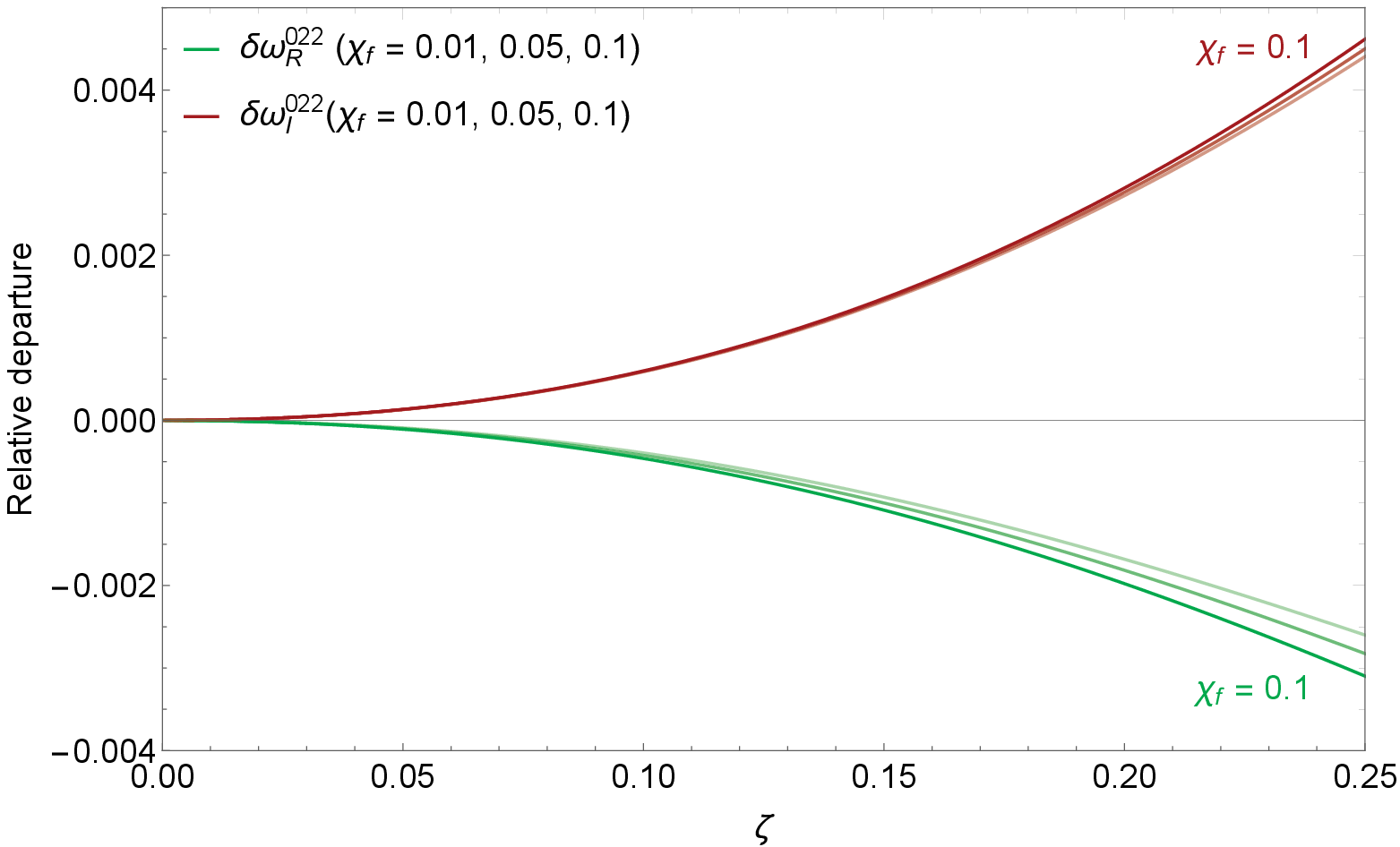}
\caption{
The dependence of the relative departures for the real and imaginary parts of QNMs with $l=m=2$ on the dimensionless deviating parameter $\zeta $.
The darker curves represent the greater spin ${\chi _f}$. 
}
\label{Fig1}
\end{figure}

The dependence of the relative departures between the QNMs of Kerr black hole on the dimensionless deviating parameter $\zeta $ for different spins ${\chi _f}$ is illustrated in Fig.~\ref{Fig1}, where $\delta \omega _{R,I}^{nlm} = \left( {\omega _{R,I}^{nlm}(\zeta ) - \omega _{R,I}^{nlm}(0)} \right)/\omega _{R,I}^{nlm}(0)$.
We note that as the spin increases, the relative departures of the real and imaginary parts of QNMs increase, which implies the correction of QNMs in EdGB gravity can be magnified by the spin.

\section{ringdown waveform and detector response} \lb{section3}

The ringdown waves from a distorted black hole consist of plus component ${h_ + }$ and cross component ${h_ \times }$, which is dominated by the form 
\begin{equation}\label{N6}
\begin{array}{l}
{h_ + }(t) = \frac{{{M}}}{{{D_L}}}\sum\limits_{l,m} {{A_{lm}}} Y_ + ^{lm}(\iota ){e^{ - t/{\tau _{lm}}}}\cos \left( {{\omega _{lm}}t - m{\phi _0} } \right),\\
{h_ \times }(t) =  - \frac{{{M}}}{{{D_L}}}\sum\limits_{l,m} {{A_{lm}}} Y_ \times ^{lm}(\iota ){e^{ - t/{\tau _{lm}}}}\sin \left( {{\omega _{lm}}t - m{\phi _0} } \right),
\end{array}
\end{equation}
for $t \ge {t_0}$ and ${h_{ + , \times }}(t) = 0$ for $t < {t_0}$, where ${t_0}$ is the initial time of ringdown.
${{M}}$ is the mass of the final black hole.
${{D_L}}$ is the luminosity distance to the source.
$l$ and $m$ are the harmonic indices.
${A_{lm}},{\phi _0} $ are the amplitude and initial phase of the QNMs.
${\tau _{lm}},{\omega _{lm}}$ are the damping time and the oscillation frequency of the QNMs determined by Eq.~(\ref{N4qnm}) in EdGB gravity.
$\iota  \in \left[ {0,\pi } \right]$ is the inclination angle of the remnant.
The function $Y_{ + , \times }^{lm}(\iota )$ indicates the total of $-2$ weighted spin spheroidal harmonics, which can be expressed as~\cite{prd-hair-loss}
\begin{equation}\label{N7}
\begin{array}{*{20}{l}}
{Y_ + ^{lm}(\iota ){ = _{ - 2}}{Y^{lm}}(\iota ,0) + {{( - 1)}^l}_{ - 2}{Y^{l - m}}(\iota ,0),}\\
{Y_ \times ^{lm}(\iota ){ = _{ - 2}}{Y^{lm}}(\iota ,0) - {{( - 1)}^l}_{ - 2}{Y^{l - m}}(\iota ,0).}
\end{array}
\end{equation}
In order to build the ringdown waveform, we focus only on two dominant modes $l = m = 2,3$ in EdGB gravity.
More specifically,
\begin{equation}\label{N8}
\begin{array}{l}
Y_ + ^{22}(\iota ) = \sqrt {\frac{5}{{4\pi }}} \frac{{1 + {{\cos }^2}\iota }}{2},Y_ \times ^{22}(\iota ) = \sqrt {\frac{5}{{4\pi }}} \cos \iota, \\
Y_ + ^{33}(\iota ) =  - \sqrt {\frac{{21}}{{8\pi }}} \frac{{1 + {{\cos }^2}\iota }}{2}\sin \iota ,Y_ \times ^{33}(\iota ) =  - \sqrt {\frac{{21}}{{8\pi }}} \sin \iota \cos \iota.
\end{array}
\end{equation}
Moreover, ${A_{lm}}$ is well fitted as~\cite{prl-ringdown-memory,prd-no-hair}
\begin{equation}\label{N8amplitude}
\begin{array}{l}
{A_{22}}(\nu ) = 0.864\nu ,\\
{A_{33}}(\nu ) = 0.44{(1 - 4\nu )^{0.45}}{A_{22}}(\nu ).
\end{array}
\end{equation}
Here $\nu  = {m_1}{m_2}/{\left( {{m_1} + {m_2}} \right)^2}$ is the symmetric mass ratio and ${m_1},{m_2}$ are masses of two separated black holes before coalescence.

\begin{figure}[htbp]
\centering
\includegraphics[scale=0.4]{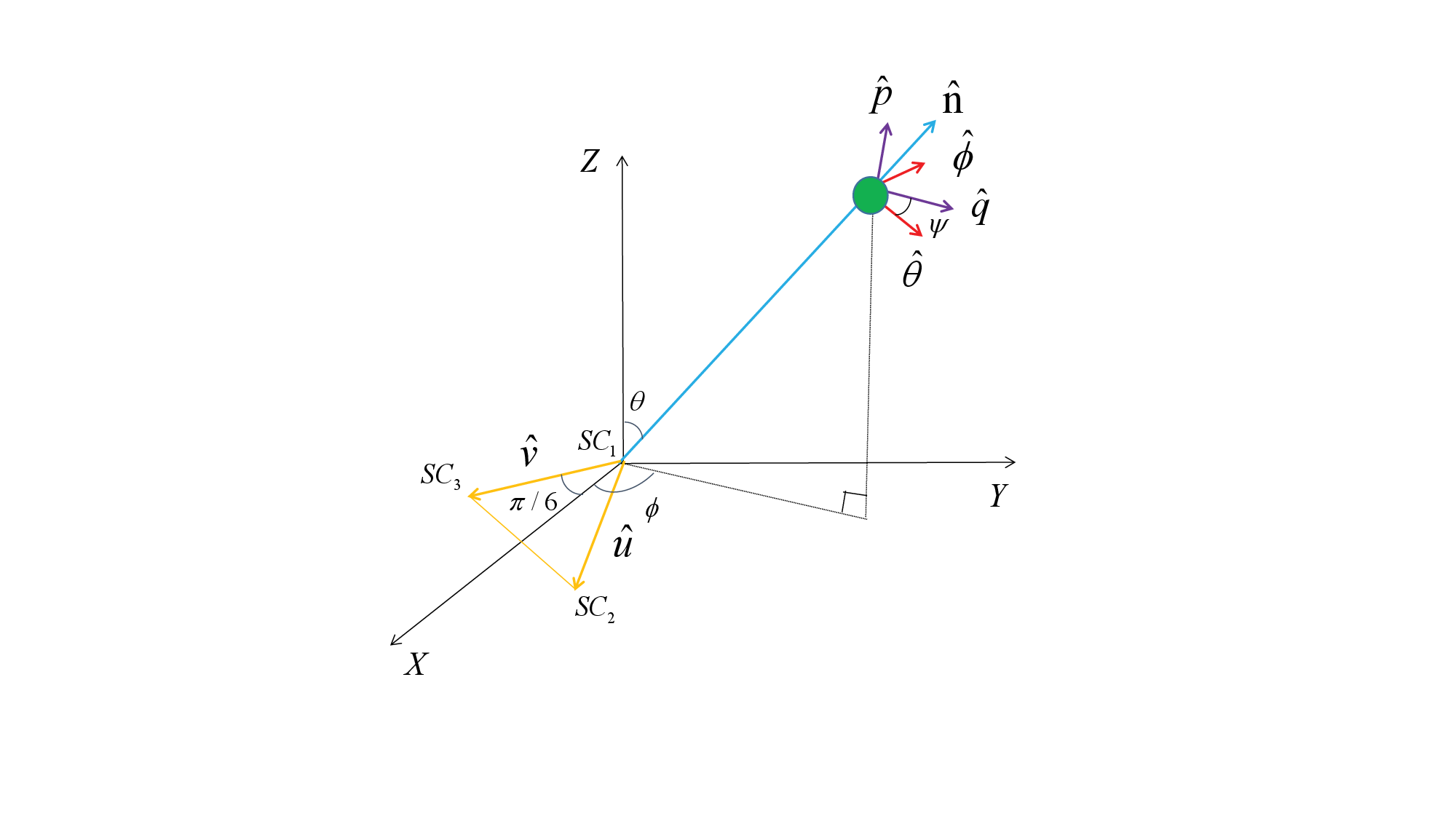}
\caption{
The detector coordinate system adopted in this paper.
}
\label{Fig2}
\end{figure}

Now, we exploit the first generation time-delay interferometry Michelson combination X~\cite{tinto2021time} to obtain the frequency-domain ringdown signals, which can be written as
\begin{equation}
h(f) = \sum\limits_{A =  + , \times } {\frac{1}{2}(1 - {e^{ - 2iu}})(D_u^A{\cal T}(u,\hat n \cdot \hat u) - D_v^A{\cal T}(u,\hat n \cdot \hat v)){h_A}(f)},
\end{equation}
where ${h_{ + , \times }}(f)$ is frequency-domain ringdown signals after the Fourier transformation of ${h_{ + , \times }}(t)$,
\begin{equation}
\begin{array}{l}
D_u^ +  = \left[ {{{\cos }^2}\theta {{\cos }^2}(\phi  - \pi /6) - {{\sin }^2}(\phi  - \pi /6)} \right]\cos 2\psi  - \cos \theta \sin (2\phi  - \pi /3)\sin 2\psi \\
D_u^ \times  =  - \cos \theta \sin (2\phi  - \pi /3)\cos 2\psi  - \left[ {{{\cos }^2}\theta {{\cos }^2}(\phi  - \pi /6) - {{\sin }^2}(\phi  - \pi /6)} \right]\sin 2\psi \\
D_v^ +  = \left[ {{{\cos }^2}\theta {{\cos }^2}(\phi  + \pi /6) - {{\sin }^2}(\phi  + \pi /6)} \right]\cos 2\psi  - \cos \theta \sin (2\phi  + \pi /3)\sin 2\psi \\
D_v^ \times  =  - \cos \theta \sin (2\phi  + \pi /3)\cos 2\psi  - \left[ {{{\cos }^2}\theta {{\cos }^2}(\phi  + \pi /6) - {{\sin }^2}(\phi  + \pi /6)} \right]\sin 2\psi. 
\end{array}
\end{equation}
The frequency-dependent transfer function ${\cal T}$ is 
\begin{equation}
{\cal T}(u,\hat n \cdot \hat u) = \frac{1}{2}{e^{ - iu}}\left[ {{e^{ - iu(1 - \hat n \cdot \hat u)/2}}{\rm{sinc}}\left( {u(1 + \hat n \cdot \hat u)/2} \right) + {e^{iu(1 + \hat n \cdot \hat u)/2}}{\rm{sinc}}\left( {u(1 - \hat n \cdot \hat u)/2} \right)} \right].
\end{equation}
Here, $u = \frac{{2\pi fL}}{c}$, where $L$ is the arm length of the detector and $c$ is the speed of light, ${\rm{sinc}}(z) = \frac{{{\rm{sin }}z}}{z}$, $\hat n = \left( {\sin \theta \cos \phi ,\sin \theta \sin \phi ,\cos \theta } \right)$ is the orientation of the source and the unit vectors with respect to detector's arms $\hat u,\hat v$ are
\begin{equation}
\hat u = \left( {\cos \frac{\pi }{6},\sin \frac{\pi }{6},0} \right),\hat v = \left( {\cos \frac{\pi }{6}, - \sin \frac{\pi }{6},0} \right).
\end{equation}

For convenience, Fig.~\ref{Fig2} shows the detector coordinate system adopted in this paper.
The origin is placed at spacecraft 1.
$(\hat p,\hat q,\hat k)$ are basis vectors of the canonical reference frame, where $\hat k$ denotes the direction of propagation of gravitational waves.
$(\hat \phi ,\hat \theta ,\hat n)$ are basis vectors of the observational reference frame and $\psi $ is the polarization angle.

In order to test EdGB gravity with space-based gravitational wave detectors, it is necessary to evaluate the capability of LISA, TaiJi, and TianQin.
Here, we adopt the noise power spectral density and average response functions of tensor polarizations for the Michelson combination X~\cite{Wang:2021owg}:  
\begin{equation}
{S_N}{(u)_X} = \frac{{4{{\sin }^2}u}}{{{u^2}}}\left[ {\frac{{s_a^2{L^2}}}{{{u^2}{c^4}}}(3 + \cos 2u) + \frac{{{u^2}s_x^2}}{{{L^2}}}} \right],
\end{equation}
\begin{equation}
\begin{array}{c}
R{(u)_X} = \frac{{{{\sin }^2}u}}{{2{u^2}}}[( - 7\sin u + 2\sin 2u)/u + ( - 4 + 5\cos u - 4\cos 2u)/{u^2} + ( - 5\sin u + 4\sin 2u)/{u^3}\\
 + (5 + \cos 2u)/3 - 6\cos 2u(Ciu - 2Ci2u + Ci3u + \ln 4/3) + 4(Ciu - Ci2u + \log 2)\\
 - 6\sin 2u(Siu - 2Si2u + Si3u)],
\end{array}
\end{equation}
where SinIntegral $\operatorname{Si}(z)=\int_0^z(\sin t / t) d t$, CosIntegral $\operatorname{Ci}(z)=-\int_z^{\infty}(\cos t / t) d t$, ${S_a}$ is the residual acceleration noise, ${S_x}$ is the displacement noise.
For LISA, ${S_a} = 3 \times {10^{ - 15}}\;{\rm{m}}{{\rm{s}}^{{\rm{ - 2}}}}\;{\rm{/}}\sqrt {{\rm{Hz}}} $, ${S_x} = 1.5 \times {10^{ - 11}}{\rm{m/}}\sqrt {{\rm{Hz}}} $ and $L = 2.5 \times {10^9}\;{\rm{m}}$~\cite{amaro2017laser}.
For TaiJi, ${S_a} = 3 \times {10^{ - 15}}\;{\rm{m}}{{\rm{s}}^{{\rm{ - 2}}}}\;{\rm{/}}\sqrt {{\rm{Hz}}} $, ${S_x} = 8 \times {10^{ - 12}}{\rm{m/}}\sqrt {{\rm{Hz}}} $ and $L = 3 \times {10^9}\;{\rm{m}}$~\cite{na-lisa-taiji}.
For TianQin, ${S_a} = 1 \times {10^{ - 15}}\;{\rm{m}}{{\rm{s}}^{{\rm{ - 2}}}}\;{\rm{/}}\sqrt {{\rm{Hz}}} $, ${S_x} = 1 \times {10^{ - 12}}{\rm{m/}}\sqrt {{\rm{Hz}}} $ and $L = \sqrt 3  \times {10^8}\;{\rm{m}}$~\cite{cqg-tianqin}.
The sky-averaged sensitivity is defined to read~\cite{cqg-lisa-sensitivity}
\begin{equation}\label{N10}
{S_n}(f) = \frac{{{S_N}(f)}}{{R(f)}}.
\end{equation}
Especially, the galactic confusion noise mainly generated by abundant double white dwarf binaries plays a non-ignorable role in the detection of gravitational waves.
For LISA and TaiJi, the galactic confusion noise takes the form~\cite{cqg-lisa-sensitivity}
\begin{equation}\label{N11}
{S_c}(f) = \alpha {f^{ - 7/3}}{e^{ - {f^\beta } + \gamma f\sin (\eta f)}}\left[ {1 + \tanh \left( {\lambda \left( {{f_c} - f} \right)} \right)} \right]{\rm{H}}{{\rm{z}}^{ - 1}}.
\end{equation}
For TianQin, the galactic confusion noise can be modeled as~\cite{prd-tianqin-galactic}
\begin{equation}
{S_{{\rm{ctq }}}}(f) = {10^{\sum\limits_{i = 0}^6 {{a_i}{{(Log(\frac{f}{{{{10}^3}}}))}^i}} {\rm{ }}}}.
\end{equation}
Provided that the detector scenario is operated for 4 years, the corresponding coefficients are $\alpha  = 9 \times {10^{ - 45}}$, $\beta  = 0.138$, $\gamma  =  - 221$, $\eta  = 521$, $\lambda  = 1680$, ${f_c} = 0.0013$, ${a_0} =  - 18.6,{a_1} =  - 1.43,{a_2} =  - 0.687,{a_3} = 0.24,{a_4} =  - 0.15,{a_5} =  - 1.8$ and ${a_6} =  - 3.2$.
Thus, the full sensitivity curve is derived by adding the galactic confusion noise to ${S_n}(f)$.

\begin{figure}[]
\centering
\includegraphics[scale=0.6]{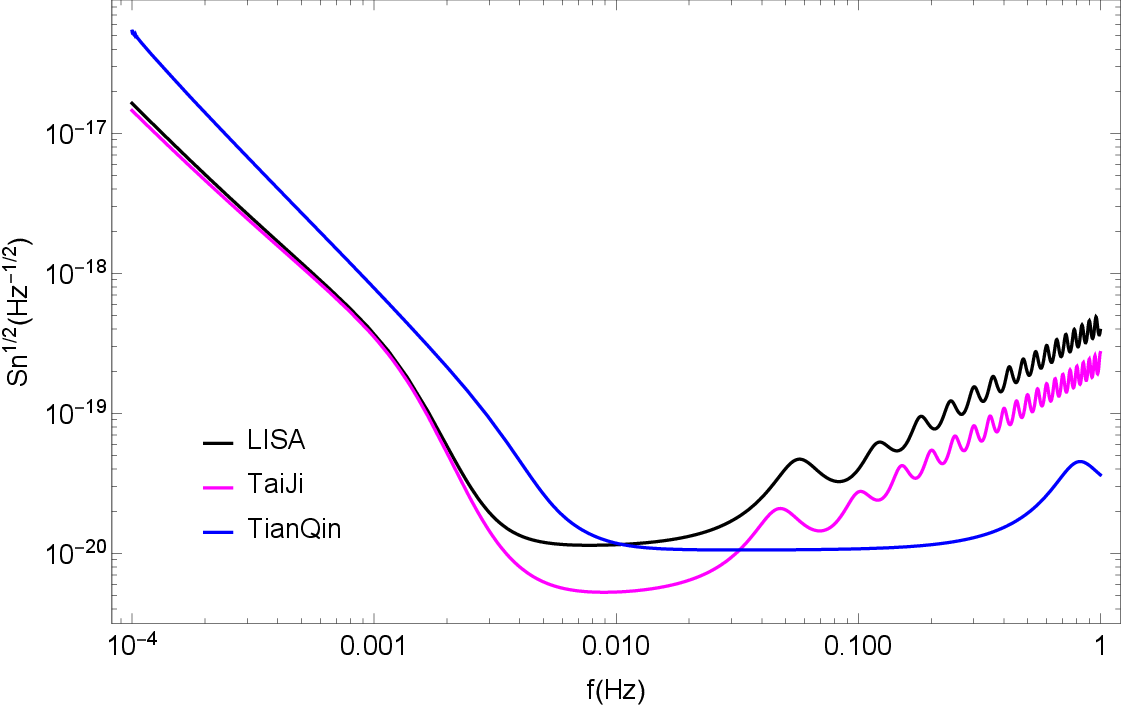}
\caption{
The sensitivity curves for LISA, TaiJi, and TianQin.
}
\label{Fig3}
\end{figure}

Fig.~\ref{Fig3} shows the sensitivity curve for LISA, TaiJi, and TianQin, from which we can see that TianQin is more sensitive to gravitational wave signals at higher frequencies while TaiJi and LISA are reliable to detect signals for lower frequencies.
TaiJi is better than LISA in detecting gravitational wave signals at higher frequencies, because the target displacement noise of TaiJi is better than LISA in case where both residual acceleration noise and arm length of the detector are similar.
Obviously, the galactic confusion noise provokes a small rise of the sensitivity value in the low-frequency range while TianQin is less affected than TaiJi and LISA.

\section{the SNR and the uncertainty of parameter estimation for EdGB gravity} \lb{section4}

The inner product weighted by the detector noise spectral density of two frequency-domain signals ${h_1}(f),{h_2}(f)$ is defined as
\begin{equation}\label{N11}
({h_1}|{h_2}) = 2\int_{{f_{low}}}^{{f_{high}}} {\frac{{{h_1}^*(f){h_2}(f) + {h_1}(f){h_2}^*(f)}}{{S_N^{}(f)}}} df,
\end{equation}
where we choose ${{f_{low}}}$ to be half of the smallest oscillation frequency and ${{f_{high}}}$ to be twice the highest oscillation frequency to prevent the ``junk'' radiation in the Fourier transformation~\cite{prd-bayesian-model}.
The sky-averaged SNR (denoted by $\rho$) based on the definition of the inner product can be expressed as
\begin{equation}\label{N11}
\rho  = \sqrt {(h|h)}.
\end{equation}
Supposing that the probability distribution for the measurement errors of parameters is Gaussian in the limit of large SNR~\cite{prd-gw-merg,prd-gw-inspiral}, 
the measurement errors on parameters ${\theta _i}$ can be derived from Fisher information matrix:
\begin{equation}\label{N11}
\Delta {\theta _i} \approx \sqrt {{{({\Gamma ^{ - 1}})}_{ii}}}.
\end{equation}
Here the Fisher information matrix is given by
\begin{equation}\label{N11}
{\Gamma _{ij}} = \left( {\frac{{\partial h}}{{\partial {\theta _i}}}\mid \frac{{\partial h}}{{\partial {\theta _j}}}} \right),
\end{equation}
where $\vec \theta $ is a 7-dimensional parameter space in the ringdown signals, namely $\vec \theta  = \left\{ {{M},{\chi _f},{D_L},\nu,\phi_0,\iota,\zeta } \right\}$.

\begin{figure}[]
\centering
\includegraphics[scale=0.6]{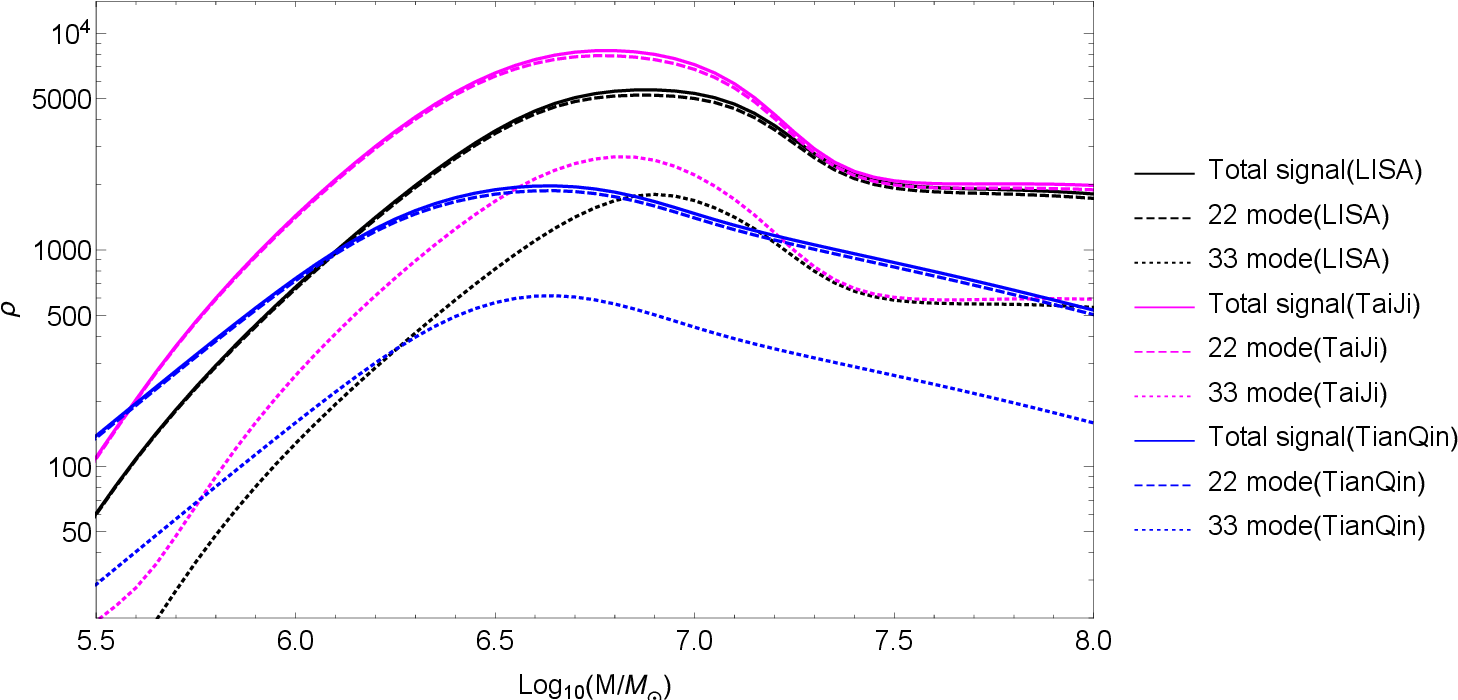}
\caption{
The SNR of LISA, TaiJi, and TianQin with the change of the mass for the ringdown signal.
The calculations are carried out with ${\chi _f} = 0.01,{D_L} = 2.5Gpc,\nu  = 2/9,{\phi _0}  = 0,\iota  = \pi /3,\zeta  = 0.2$.
}
\label{Fig4}
\end{figure}

First, the SNR varying with the mass of the black hole for LISA, TaiJi, and TianQin are calculated in Fig.~\ref{Fig4}, from which we plot two dominant QNMs in the ringdown signals, respectively.
As one can see, the total SNR is heavily dominated by the strongest $(2,2)$ mode.
On the whole, as the mass increase, the SNR grows until reaching the maximum and then decreases gradually for all detectors.
After comparing the SNR of LISA, TaiJi, and TianQin, it is found that TianQin is more sensitive to gravitational signals with smaller masses, while TaiJi and LISA is more reliable to detect signals for more massive black holes.
Especially, the galactic confusion noise serves as a catalyst for a small dip to appear around the mass within $6 \times {10^6}{M_ \odot } \mathbin{\lower.3ex\hbox{$\buildrel<\over{\smash{\scriptstyle\sim}\vphantom{_x}}$}} M \mathbin{\lower.3ex\hbox{$\buildrel<\over{\smash{\scriptstyle\sim}\vphantom{_x}}$}} {10^8}{M_ \odot }$ for TaiJi and LISA, while for TianQin, this effect is trivial.
Besides, its implications can be negligible for the less massive black hole.

\begin{figure}[]
\centering
\includegraphics[scale=0.58]{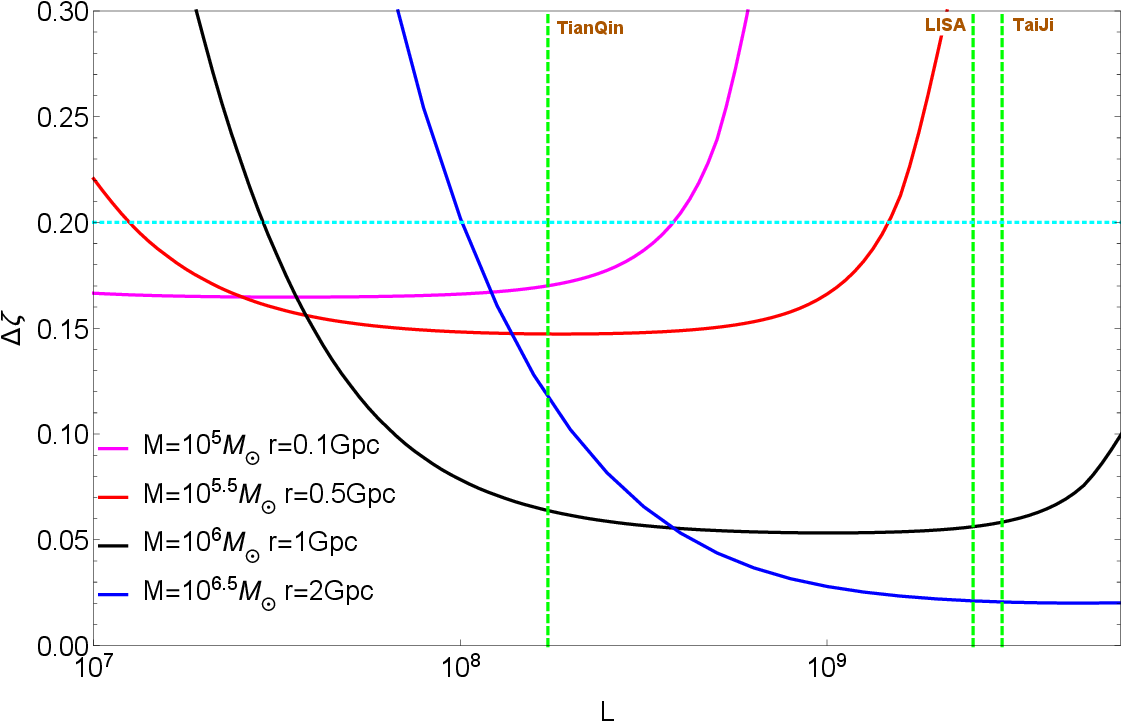}
\caption{
The dependence of parameter estimation accuracy $\Delta \zeta $ on the arm length of the detector $L$ for different sources of gravitational waves.
Here, we choose the basic parameters of LISA as a reference.
The residual acceleration noise is fixed as ${S_a} = 3 \times {10^{ - 15}}\;m{s^{ - 2}}\;/\sqrt {Hz} $.
The displacement noise is proportional to the arm length ${S_x} \sim \alpha L$, where $\alpha  = \frac{{1.5 \times {{10}^{ - 11}}}}{{2.5 \times {{10}^9}}}$.
The dotted cyan horizontal line represents the maximum error of $\zeta $ and the dashed green vertical line denotes the real arm length of LISA, TaiJi, and TianQin.
The calculations are carried out with ${\chi _f} = 0.01,\nu  = 2/9,{\phi _0}  = 0,\iota  = \pi /3,\zeta  = 0.2$.
}
\label{Fig5}
\end{figure}

Next, we consider estimating the measurement errors for the dimensionless deviating parameter $\zeta $ (denoted by $\Delta \zeta $) via available tools.
Before introducing the standard parameters of the space-based gravitational wave detector, we roughly evaluate the influence of the arm length of the detector on the dimensionless deviating parameter, where the residual acceleration noise is fixed and the displacement noise is proportional to the arm length for gravitational wave detection.
Fig.~\ref{Fig5} shows the measurement errors as a function of arm length for different sources of gravitational waves, where the maximum error of $\zeta $ is denoted with dotted cyan horizontal line and the real arm length of LISA, TaiJi, and TianQin is denoted with dashed green vertical line.
The test of EdGB gravity will be affected by sources of different masses. 
The arm length of TaiJi and LISA is more appropriate to test EdGB gravity for more massive black holes while for less massive black holes, more practical arm length can be provided by TianQin.
Because the above are qualitative analysis, only the parameters related to LISA are used here to evaluate the ability to test EdGB gravity on the whole.
For the specific calculation below, the standard parameters of the space-based gravitational wave detector are considered.

\begin{figure}[htbp]
\centering
\includegraphics[width=0.4\textwidth]{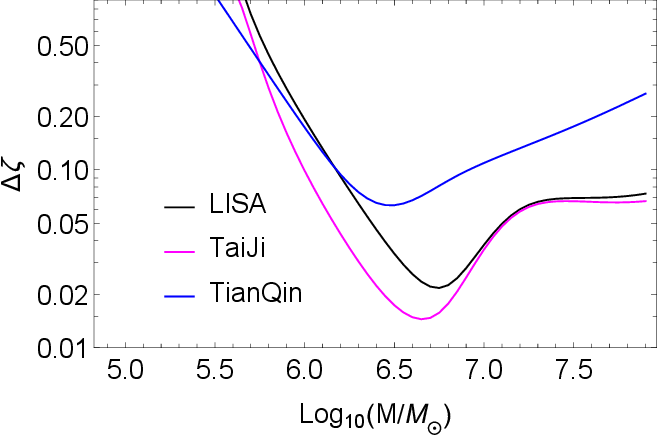}
\includegraphics[width=0.399\textwidth]{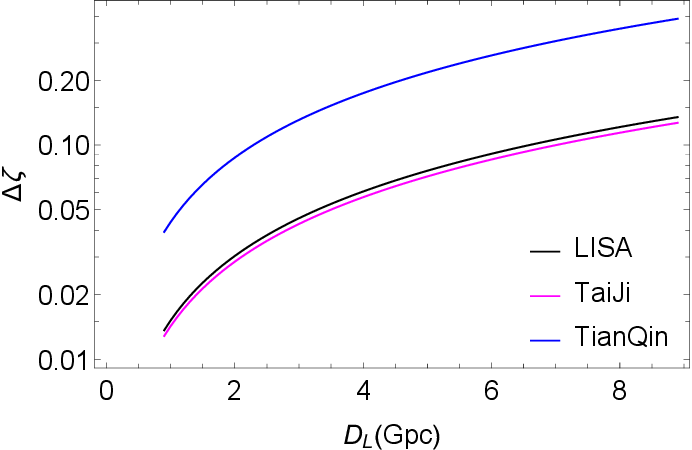}\\
\includegraphics[width=0.411\textwidth]{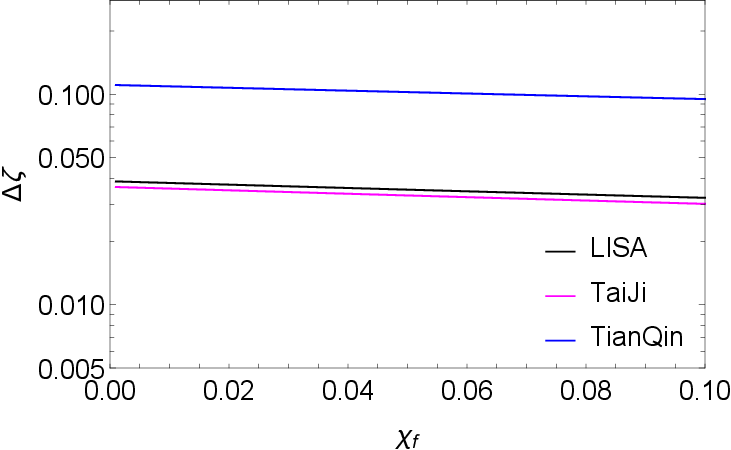}
\includegraphics[width=0.411\textwidth]{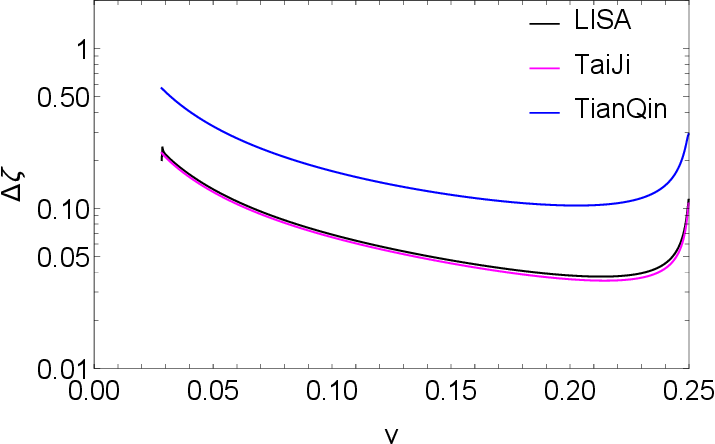}
\caption{
The dependence of parameter estimation accuracy $\Delta \zeta $ on the mass $M$ (top left), the luminosity distance ${D_L}$ (top right), the spin of the remnant black hole ${{\chi _f}}$ (bottom left) and the symmetric mass ratio $\nu $ (bottom right).
The black, magenta, and blue curves represent the errors detected by LISA, TaiJi, and TianQin respectively.
Other parameters used are ${\phi _0}=0,\iota  = \pi /3,\zeta  = 0.2$.
}
\label{Fig6}
\end{figure}

In addition, in order to explore how several source parameters, such as the mass ${{M}}$, the luminosity distance ${D_L}$, the spin of the remnant black hole ${{\chi _f}}$ and the symmetric mass ratio $\nu $, affect $\Delta \zeta $, we display the variation of $\Delta \zeta $ with related parameters in Fig.~\ref{Fig6}.
The top left plot of Fig.~\ref{Fig6} shows the measurement errors as a function of the mass, where we have assumed ${\chi _f} = 0.01,{D_L} = 2.5Gpc,\nu  = 2/9$.
As expected, at first $\Delta \zeta $ decreases with the increase of ${{M}}$ until arriving at the minimum value, but soon increases with the accumulation of ${{M}}$ for all detectors.
For specific massive black holes, a slight bulge emerges in the measurement errors owing to the galactic confusion noise, which implies the galactic confusion noise plays a negative role in constraining EdGB gravity.
It is clear that TianQin can constrain $\zeta $ more accurately for smaller masses, but for more massive black holes TaiJi and LISA performs well, which is also reflected by the sensitivity of detectors.
The measurement errors as a function of the luminosity distance are plotted in the top right plot of Fig.~\ref{Fig6}, where we fix $M ={10^7}{M_ \odot },{\chi _f} = 0.01,\nu  = 2/9$. 
As one can see, $\Delta \zeta $ increases with the increase of ${D_L}$, which is obvious because the SNR increases monotonically with decreasing distance.
This is intuitively reflected in the ringdown waveform containing the term $1/{D_L}$, so there is a $1/{D_L}^2$ in the error function after Fisher analysis.
The bottom left plot of Fig.~\ref{Fig6} presents the dependence of measurement errors on the spin of the remnant black hole, where we set $M ={10^7}{M_ \odot },{D_L} = 2.5Gpc,\nu  = 2/9$.
We note that $\Delta \zeta $ decreases extremely slowly as ${{\chi _f}}$ increases.
It was implied that the spin parameter of the black hole has little effect on measurement errors for EdGB gravity by detectors.
The dependence of measurement errors on the symmetric mass ratio is illustrated in the bottom right plot of Fig.~\ref{Fig6}, where we choose $M ={10^7}{M_ \odot },{D_L} = 2.5Gpc,{\chi _f} = 0.01$. 
Observe first that $\Delta \zeta $ decrease with the increase of $\nu $ and then grows larger abruptly as $\nu $ approaches $0.25$.
That's because radiated energy gets more with large symmetric mass ratio and as $\nu $ approaches $0.25$, the amplitude of the QNMs for $(3,3)$ is zero.
In order to estimate associated parameters better, it was necessary to avoid selecting $\nu $ in this limit or replace it with other dominant modes.

\begin{figure}[]
\centering
\includegraphics[scale=0.6]{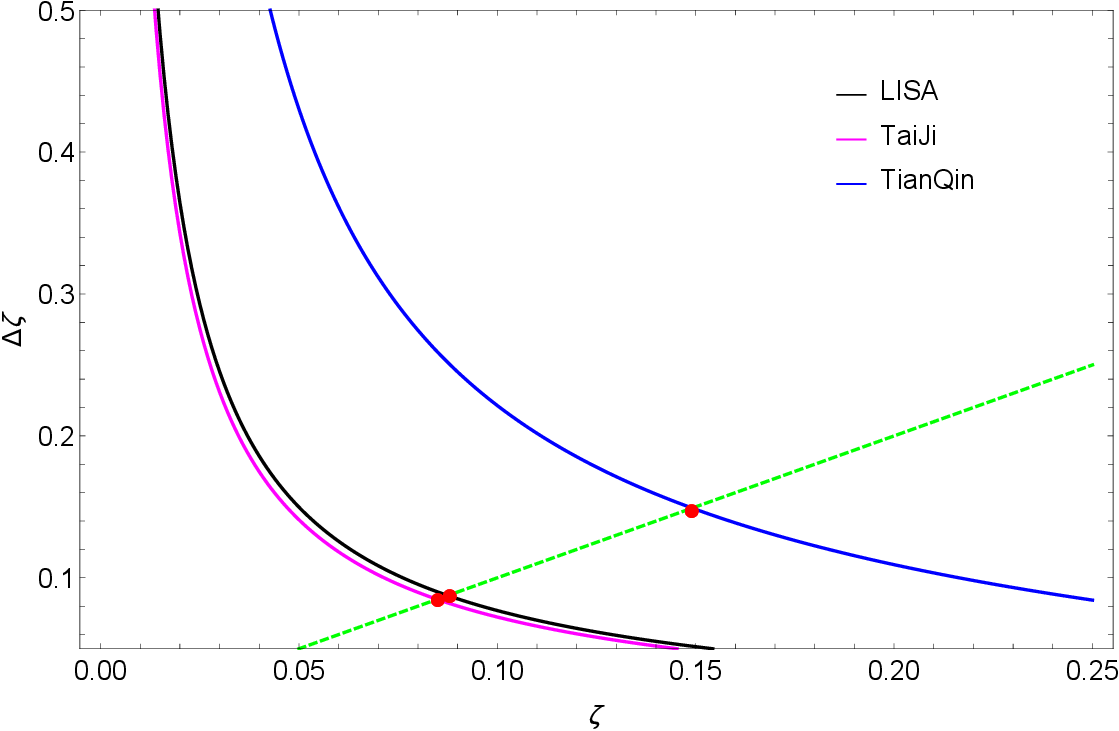}
\caption{
The dependence of parameter estimation accuracy $\Delta \zeta $ on the dimensionless deviating parameter $\zeta $.
The dashed green line denotes $\Delta \zeta = \zeta$.
The red intersections are $\Delta {\zeta _{\max }}$ for each detector.
The profiles are obtained with given $M = {10^7}{M_ \odot },{\chi _f} = 0.01,{D_L} = 2.5Gpc,\nu  = 2/9,{\phi _0}  = 0,\iota  = \pi /3$ for different space-based gravitational wave detectors.
}
\label{Fig7}
\end{figure}

\begin{figure}[htbp]
\centering
\includegraphics[scale=0.5]{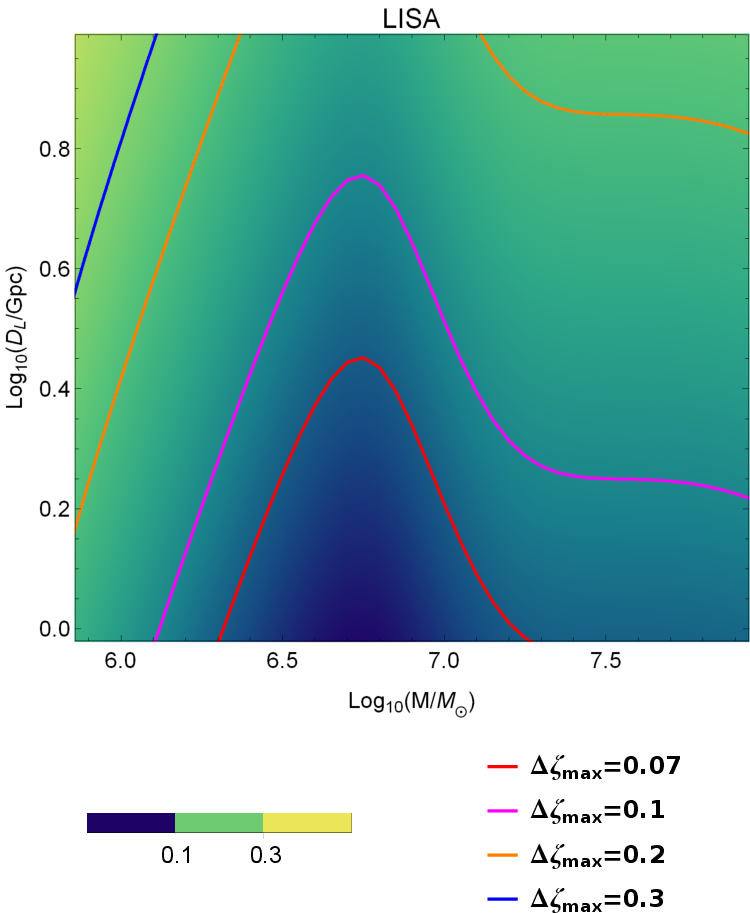}
\includegraphics[scale=0.5]{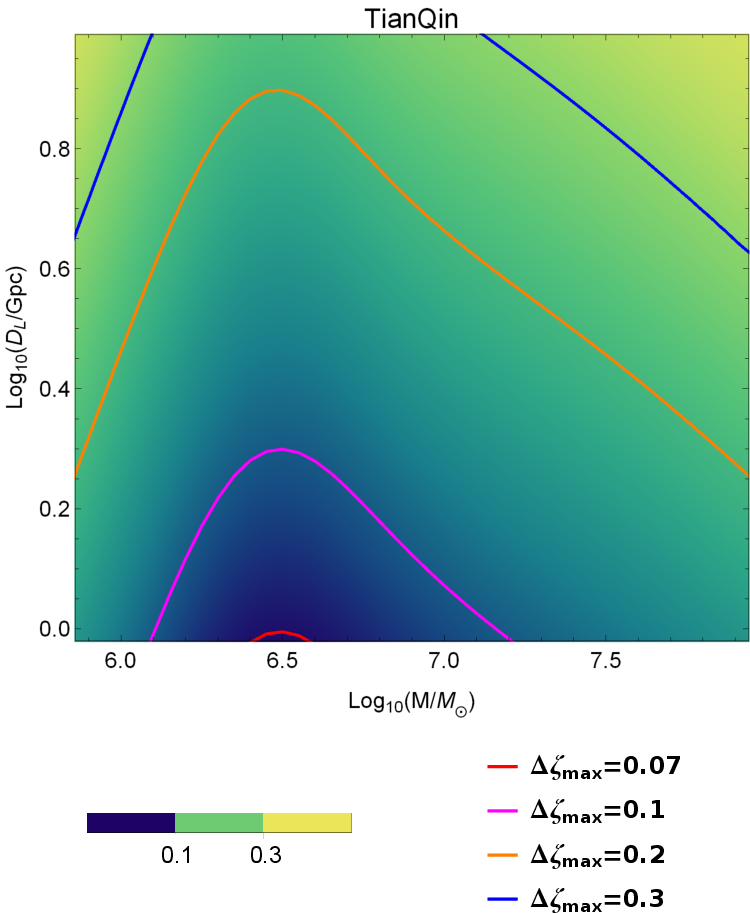}
\caption{
The maximum constraint on the dimensionless deviating parameter for EdGB gravity with LISA and TianQin in the $Log({D_L}/Gpc) - Log({M}/{M_ \odot })$ plane.
The calculations are carried out for the case of ${\phi _0}  = 0,\iota  = \pi /3,\nu  = 2/9,{\chi _f} = 0.01$.
}
\label{Fig8}
\end{figure}
 
\begin{figure}[htbp]
\centering
\includegraphics[scale=0.61]{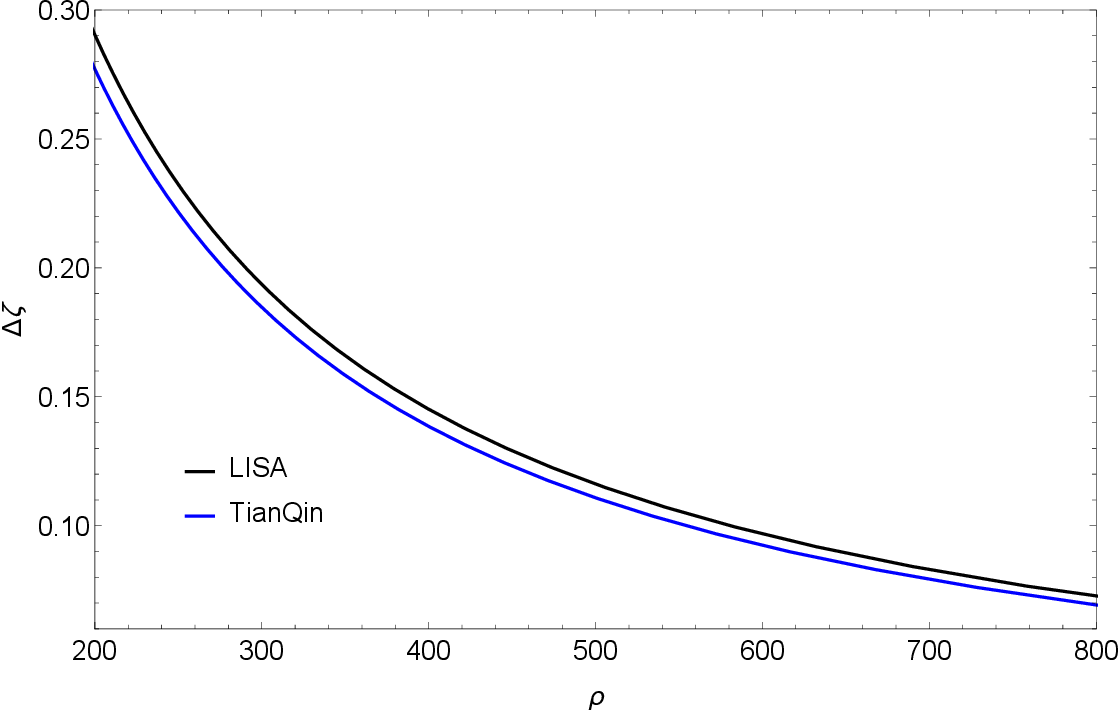}
\caption{
The dependence of parameter estimation accuracy $\Delta \zeta $ on the SNR.
The calculations are carried out with ${M} = {10^6}{M_ \odot },{\chi _f} = 0.01,\nu  = 2/9,{\phi _0}  = 0,\iota  = \pi /3,\zeta  = 0.2$ for LISA and TianQin.
}
\label{Fig9}
\end{figure}

In particular, we pay attention with interest to the issue of the measurement errors $\Delta \zeta $ varying with $\zeta$, which is illustrated in Fig.~\ref{Fig7}.
One can see that the effect of $\Delta \zeta $ on $\zeta$ is dynamic.
That's because $\zeta$ is nonlinear in Eq.~(\ref{N4qnm}) and Eq.~(\ref{N5qnm2}).
Owing to relatively large coefficients in Table~\ref{tab1}, higher-order corrections are not discarded here.
Hence, the result of the covariance matrix contains variable $\zeta$, and specifically, $\Delta \zeta $ decreases with the increase of $\zeta$.
When the dimensionless deviating parameter is small, it seems difficult to distinguish EdGB gravity from General Relativity due to the larger uncertainty of $\zeta$.
We emphasize this problem by tracing the maximum error of $\zeta$ in EdGB gravity (denoted by $\Delta {\zeta _{\max }}$), namely the solution of the equation $\Delta {\zeta } = \zeta$, which is treated as the maximum constraint of the detector by probing a particular wave source, marked as red dots.
Only in the area below that do the space-based gravitational wave detectors have the potential to tell the difference between EdGB gravity and General Relativity.
Due to no significant difference in the standard parameters of the space-based gravitational wave detector between TaiJi and LISA, for comparison purposes, the remainder of our paper focuses only on the results of LISA and TianQin.
Furthermore, to investigate the maximum constraint on the dimensionless deviating parameter for EdGB gravity with LISA and TianQin, we display the density plots of $\Delta {\zeta _{\max }}$ in the $Lo{g_{10}}({D_L}/Gpc) - Lo{g_{10}}({M}/{M_ \odot })$ plane in Fig.~\ref{Fig8}.
We observe that $\zeta $ can be best constrainted with LISA for ${M} \sim 5.5 \times {10^6}{M_ \odot }$ and TianQin for ${M} \sim 3 \times {10^6}{M_ \odot }$.
What's more, $\Delta \zeta $ decreases with the increase of SNR, which is demonstrated in Fig.~\ref{Fig9}.
For a much smaller deviation from General Relativity, we need to count on detectors with rather larger SNR to obtain the maximum constraint.
The growth of the required SNR is not linear, and as the accuracy of constraints increases, the SNR increases dramatically. 
It was noticeable that compared with TianQin, LISA is more likely to give an optimal constraint in the future.

\section{Bayesian inference} \lb{section5}

To verify the results calculated by Fisher information matrix analysis, we employ Bayesian inference method to analyze the simulated source.
Bayesian inference method is widely used in estimating the probability distribution of unknown parameters from sampled data containing signals and noise, which is instrumental in astrophysical and cosmological analysis.
Unlike Fisher information matrix analysis, which is limited to large SNR, Bayesian analysis is applicable to a wider range of situations. 
Furthermore, Bayesian posterior probability distributions will give more information.
The disadvantage is that it is more computationally intensive. 
In essence, Bayesian statistics is to create a likelihood to associate unknown parameters and measurement parameters.
Then the probability distribution of the unknown parameters will be updated through the distribution of measurement data.
This process is based on Bayes'theorem:
\begin{equation}\label{N11}
P(\vec \theta |d) = \frac{{\pi (\vec \theta ) \mathcal{L}(d|\vec \theta )}}{{p(d)}},
\end{equation}
where $P(\vec \theta |d)$ is the posterior probability of the set of free parameters.
$d = h({\vec\theta _0}) + n$ represents the measurement data which collects gravitational-wave signal modeled by all the true parameters $\vec\theta _0$ and detector noise $n$ modeled by the noise power spectra.
${\pi (\vec \theta )}$ is the prior probability of $\vec \theta $.
${p(d)}$ is a normalization constant which is also called the evidence of $d$.
${\mathcal{L}(d|\vec \theta )}$ is the likelihood, which can be written as
\begin{equation}\label{N11}
\mathcal{L}(d|\vec \theta ) = \exp \left[ { - \frac{1}{2}(h(\vec \theta ) - d|h(\vec \theta ) - d)} \right].
\end{equation}

\begin{figure}[htbp]
\centering
\includegraphics[scale=0.6]{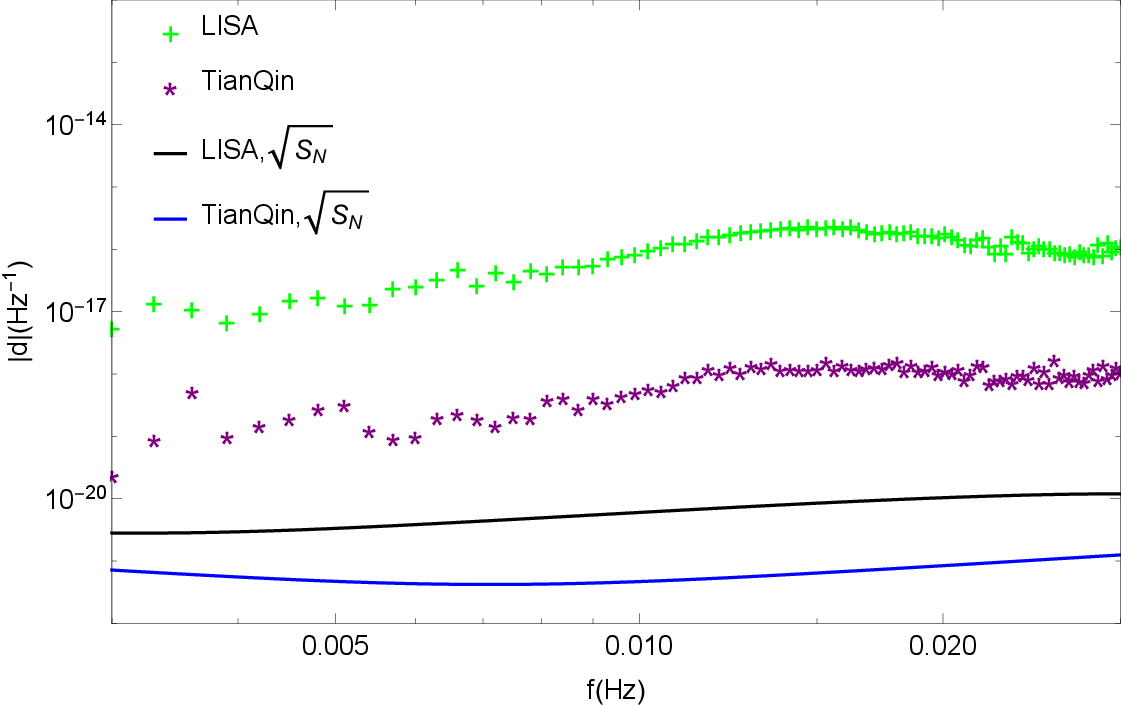}
\caption{
The amplitudes of the strain data $d$ for ringdown and the noise power spectra for LISA and TianQin.
The results are obtained using the parameters $M = 6 \times {10^6}{M_ \odot },{\chi _f} = 0.01,{D_L} = 2.5Gpc,\nu  = 2/9,{\phi _0}  = 0,\iota  = \pi /3,\zeta  = 0.2$.
}
\label{Fig10}
\end{figure}

The amplitudes of the strain data for ringdown and the noise power spectra for LISA and TianQin are illustrated in Fig.~\ref{Fig10}.
As one can see that the highest peak corresponds to mode $(2,2)$. 
By comparison, the noise power spectra of TianQin reaches the lowest level in the present frequency band.

\begin{figure}[htbp]
\centering
\includegraphics[scale=0.35]{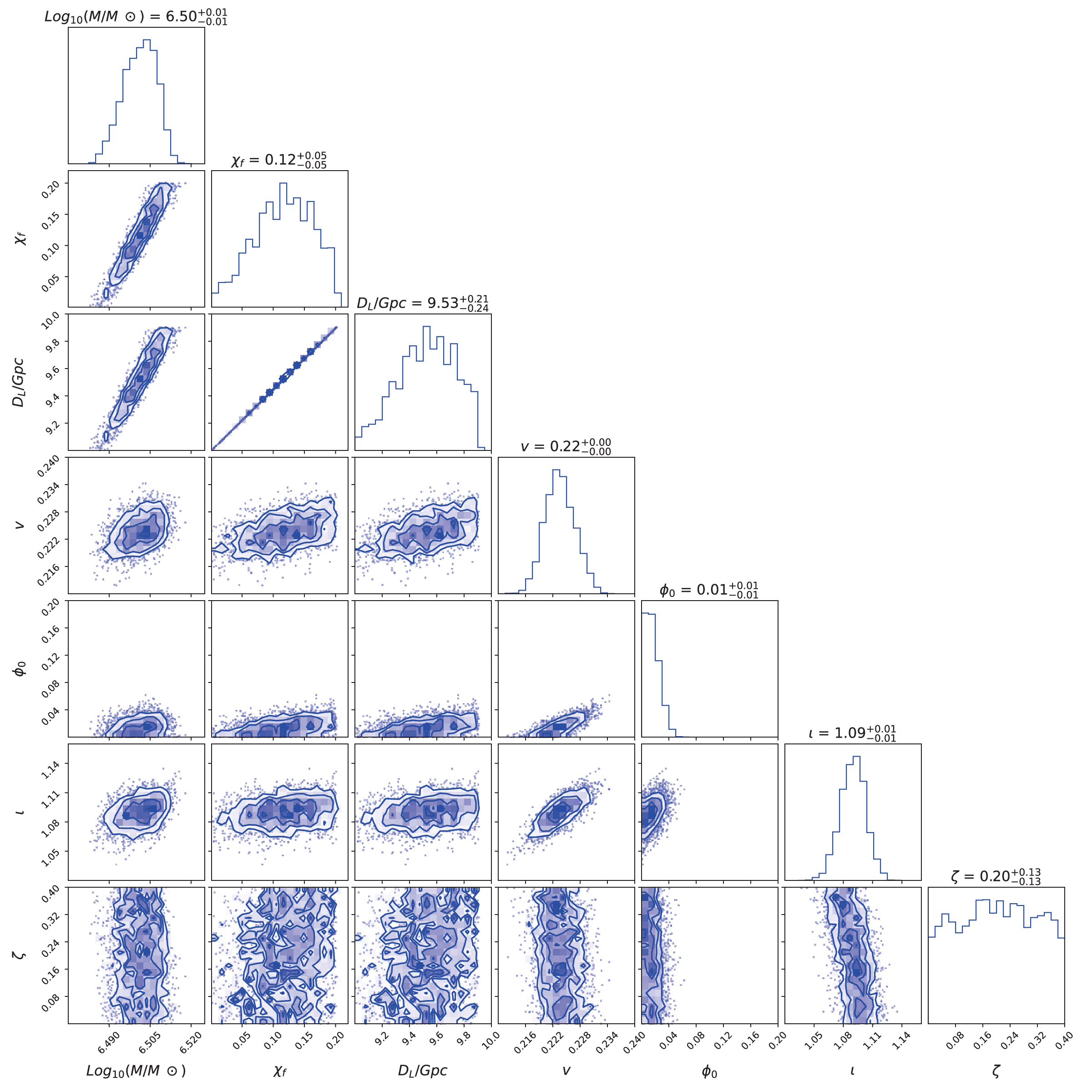}
\caption{
The posterior distribution for the ringdown signals with LISA.
The true parameters are set with ${M} = {10^{6.5}}{M_ \odot },{\chi _f} = 0.1,{D_L} = 10Gpc,\nu  = 2/9,{\phi _0}  = 0,\iota  = \pi /3,\zeta  = 0.2$.
}
\label{Fig11}
\end{figure}

\begin{figure}[htbp]
\centering
\includegraphics[scale=0.35]{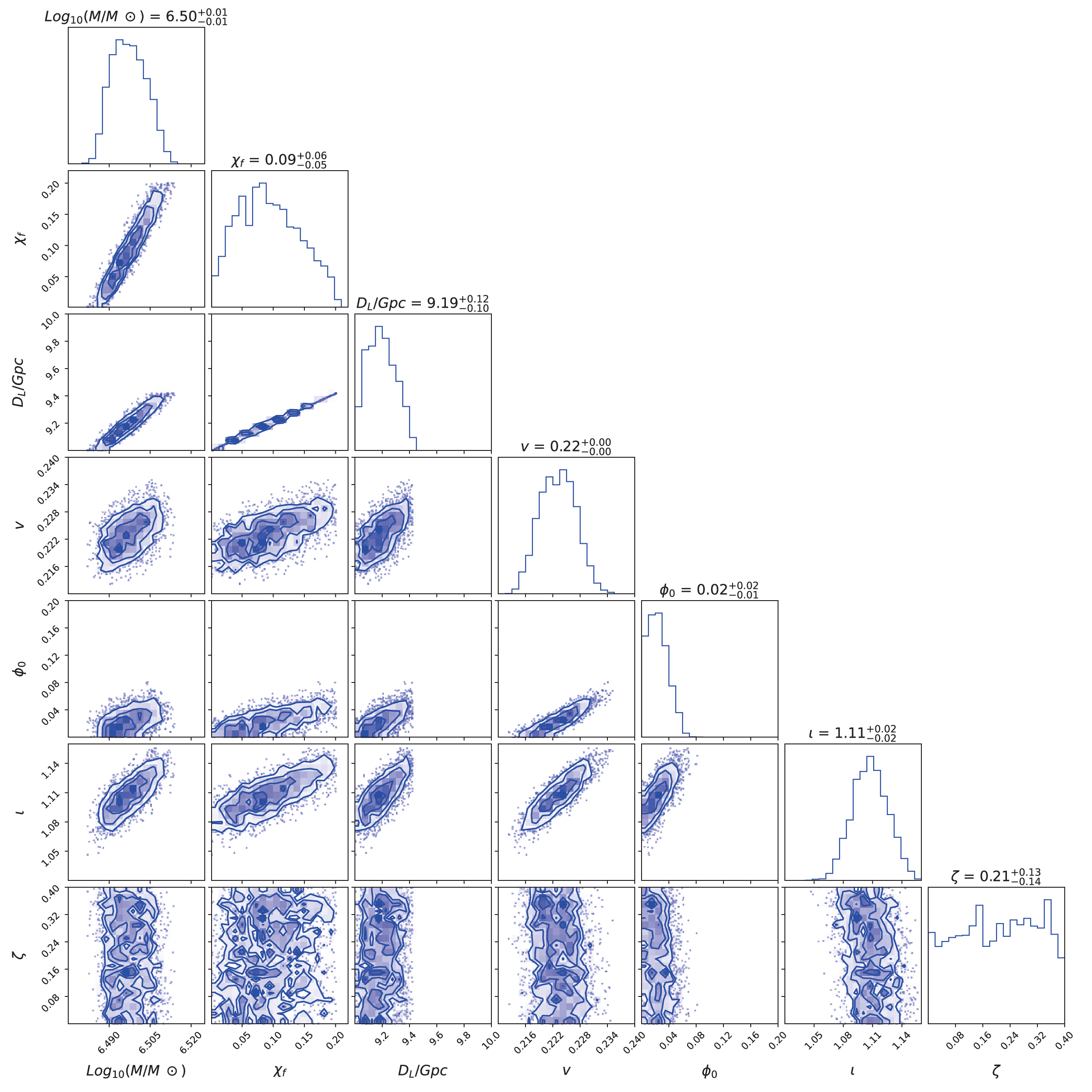}
\caption{
The posterior distribution for the ringdown signals with TianQin.
The true parameters are set with ${M} ={10^{6.5}}{M_ \odot },{\chi _f} = 0.1,{D_L} = 10Gpc,\nu  = 2/9,{\phi _0}  = 0,\iota  = \pi /3,\zeta  = 0.2$.
}
\label{Fig12}
\end{figure}

\begin{figure}[htbp]
\centering
\includegraphics[scale=0.5]{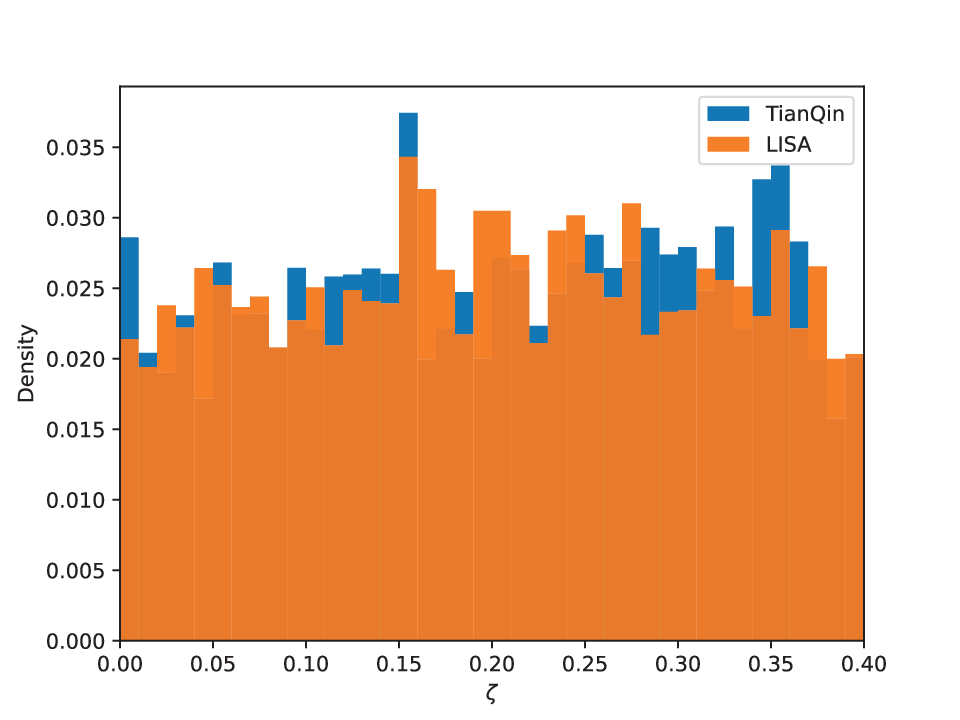}
\includegraphics[scale=0.5]{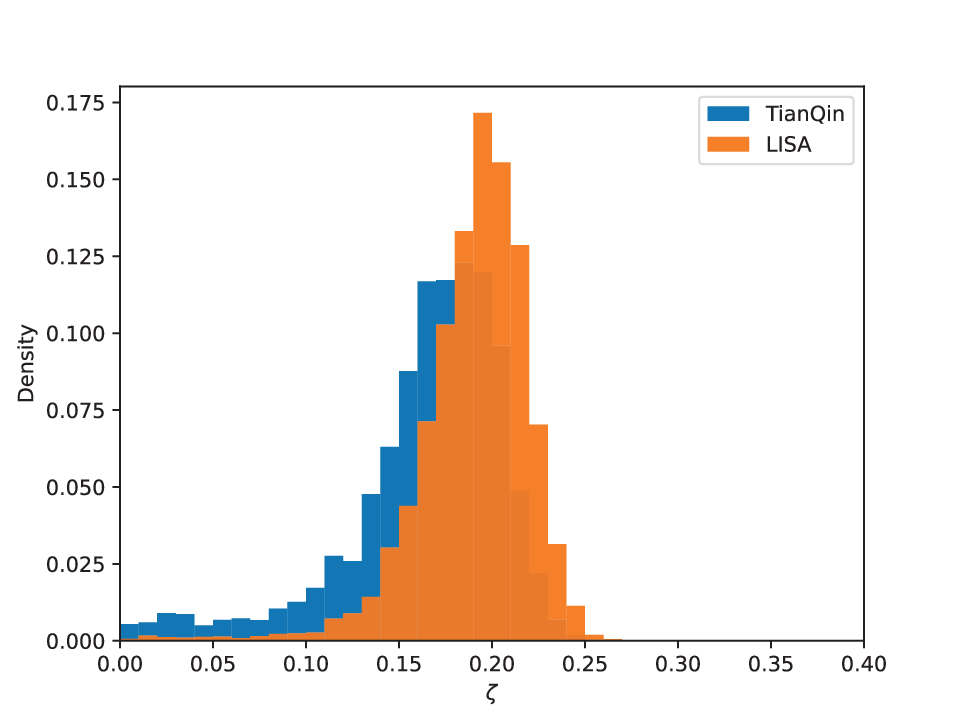}
\caption{
The posterior distribution of $\zeta $ for LISA and TianQin.
The true parameters are set with ${M} = {10^{6.5}}{M_ \odot },{\chi _f} = 0.1,\nu  = 2/9,{\phi _0 }  = 0,\iota  = \pi /3,\zeta  = 0.2$.
We set ${D_L} = 10Gpc$ for the left plot and ${D_L} = 0.5Gpc$ for the right plot.
}
\label{Fig13}
\end{figure}

After generating simulation data, we utilize Bayesian inference method to obtain the probability distributions of the source parameters, including $Lo{g_{10}}({M}/{M_ \odot })$, ${\chi _f}$, ${D_L}/Gpc$, $\nu$, ${\phi _0 }$, $\iota$ and $\zeta $. 
Fig.~\ref{Fig11} shows the posterior distribution for the ringdown signal with LISA, where the true parameters for the ringdown signal are set to be ${M} = {10^{6.5}}{M_ \odot },{\chi _f} = 0.1,{D_L} = 10Gpc,\nu  = 2/9,{\phi _0 }=0,\iota  = \pi /3,\zeta  = 0.2$. 
The priors of the corresponding parameters are respectively set to be uniform distributions within the range of $[6,7]$, $[0.001,0.2]$, $[9,14]$, $[0,1/4]$, $[0,2\pi ]$, $[0,\pi ]$, $[0,0.4]$.
For comparison, the results for the same wave source with TianQin are shown in Fig.~\ref{Fig12}.
As we can see that the probability distribution for ${D_L}$ is relatively poor, even if we have set a narrower prior.
If a broad priori is set, the estimation accuracy of ${D_L}$ will be even worse.
More seriously, its estimation will affect the estimation accuracy of other parameters.
Fortunately, the ringdown signal is usually spotted after inspiral and merger, which will provide an estimate of the luminosity distance.
Now that we focus only on the probability distribution for $\zeta $, Fig.~\ref{Fig13} displays the posterior possibility of $\zeta $ for LISA and TianQin with different luminosity distances to the source.
As shown in the left side of Fig.~\ref{Fig13}, the luminosity distance is $10Gpc$, and the dimensionless deviating parameter cannot be distinguished from 0.
The 95\% credible upper limits given by different detectors are 0.385 for LISA and 0.387 for TianQin.
For the right subplot, all parameters remain the same, except that the luminosity distance is reduced to $0.5Gpc$.
For such a signal with an extremely large SNR (over 1000), the space-based detectors can distinguish between dimensionless deviating parameter and 0 by the ringdown signal.
The estimates for the dimensionless deviating parameter with both detectors are $0.1943^{+0.0287}_{-0.0216}$ for LISA and $0.1727^{+ 0.0402}_{- 0.0292}$ for TianQin. 
It is obvious that LISA can give the best limit.
It is found that the posterior possibility of $\zeta$ at shorter luminosity distances is better than that at longer luminosity distance, which is implied that possible constraint to $\zeta $ become more accurate at high SNR. 
The uncertainty of the dimensionless deviating parameter for different detectors matches the results of the Fisher information matrix.

\section{Concluding remarks} \lb{section6}

In this paper, we analyze the ability of space-based gravitational-wave detectors LISA and TianQin to constrain the dimensionless deviating parameter for EdGB gravity by detecting ringdown signals.
The detection capabilities of different detectors for ringdown signals are first evaluated and compared.
The ringdown waveform is parameterized by several of the strongest modes in EdGB gravity.
We adopt time-delay interferometry Michelson combination X to make scientific performance evaluations for space-based detectors.
According to the SNR distribution of LISA, TaiJi, and TianQin, it is found that TianQin is more sensitive to the ringdown signals with the less massive black hole, while TaiJi and LISA are more reliable to detect signals for more massive black holes.
For specific massive black holes, the effect of galactic confusion noise plays an important role in the detection signal, and this effect is insignificant for TianQin compared to TaiJi and LISA.

In order to estimate the measurement accuracy of the dimensionless deviating parameter, we first used Fisher information matrix analysis to study qualitatively the influence of the arm length of the detector on the measurement errors and then explore how the constraints on the dimensionless deviating parameter are affected by the source parameters, such as the mass, the luminosity distance, the spin of the remnant black hole and the symmetric mass ratio.
In particular, we have found that the measurement errors of the dimensionless deviating parameter increase as the dimensionless deviating parameter decreases.
To distinguish between EdGB gravity and General Relativity, it is necessary to determine whether the dimensionless deviating parameter is 0, which requires that its uncertainty is less than the value of the parameter itself. 
For a given source, the critical value at which the uncertainty is equal to the dimensionless deviating parameter is the upper limit of the detector's ability to constrain the dimensionless deviating parameter for EdGB gravity.
We present the maximum capability of the detectors to constrain the dimensionless deviating parameter with the distribution of the wave source.
LISA has more potential to constrain the dimensionless deviating parameter to an accurate level for massive binary black hole mergers, while TianQin performs better for smaller black holes.

In addition, to verify the conclusion obtained through Fisher information matrix analysis, we have performed Bayesian parameter estimation on the simulated data.
It is found that the posterior possibility of $\zeta $ becomes better as luminosity distances decrease. 
By comparison, LISA might be better able to constrain the dimensionless deviating parameter for massive black holes, because the sensitivity in this frequency band is lower than that of other detectors. 
These results may be helpful in testing EdGB gravity with future space-based gravitational wave detectors.
For a more realistic black hole with bigger spins, it is interesting to generalize the research to explore SNR and estimate deviating parameters.

\section*{Acknowledgments}
This work is supported by the National Key ${\rm{R\& D}}$ Program of China under Grant No.2022YFC2204602, the Natural Science Foundation of China Grant No.11925503.

\bibliographystyle{h-physrev}
\bibliography{references_shao}

\end{document}